\documentstyle[sprocl,epsfig,wrapfig]{article}

\def\be{\begin{equation}}
\def\ee{\end{equation}}
\def\bea{\begin{eqnarray}}
\def\eea{\end{eqnarray}}
\def\pvec{{$\vec {\rm p}$}} %p with an arrow on top

\def\lesim{{$\lower 2pt\hbox{$\scriptstyle <$} 
\atop\raise4pt\hbox{$\scriptstyle\sim$}$}} %%less than with a \sim under it
 
\def\grsim{{$\lower2pt\hbox{$\scriptstyle >$} \atop\raise4pt\hbox
{$\scriptstyle\sim$}$}}
\begin{document}

\title{THE RHIC SPIN PROGRAM: SNAPSHOTS OF PROGRESS}

\author{S. E. VIGDOR}

\address{Dept. of Physics, Indiana University and Indiana University
Cyclotron Facility,\\Bloomington IN 47405, USA\\E-mail: 
vigdor@iucf.indiana.edu} 


\maketitle\abstracts{I review progress toward the experimental study of
polarized proton collisions at RHIC, at center-of-mass energies of
several hundred GeV.  The tools under development for these experiments
are summarized, with emphasis on the complementarity for the spin
program of the two major detectors, PHENIX and STAR.  The proposed
research program includes measurements of the spin structure of hadrons,
tests of QCD predictions for spin observables, and polarization searches
for interactions beyond the Standard Model.  I argue, in particular, that
RHIC should provide the best determination of the gluonic contribution
to proton spin foreseen for the coming decade.}

\section{Introduction}

About one year from today, the Relativistic Heavy Ion Collider (RHIC)
at Brookhaven will commence operation, primarily to study the formation
and decay of strongly interacting matter at extremely high energy densities.
The availability of polarized proton beams at RHIC will facilitate a
complementary program of hadronic spin studies at unprecedentedly high
energies ($\sqrt{s} = 50 - 500$ GeV) and momentum transfers ($p_T$\grsim
10 GeV/c), where low-order perturbative QCD (pQCD) should be viable.
The projected luminosities for colliding polarized beams are sufficiently
high (up to $2 \times 10^{32}$ cm$^{-2}$s$^{-1}$) to permit investigation
of relatively rare processes.  This new capability will enable a program
of unique and definitive experiments to: (1) measure the spin structure
of hadrons, especially the contributions from gluons and sea antiquarks;
(2) search for physics beyond the Standard Model, {\em  e.g.}, via
anomalous parity violation in very hard hadronic collisions; and
(3) test pQCD predictions for polarization observables.  The RIKEN
institute in Japan has provided much of the foresight and generous
funding contributions that make this entire spin program possible.

Overviews of the projected RHIC spin program have been provided in
talks at previous symposia in this series.\cite{{Ma96},{En96}}  
In the present talk, I would therefore like to highlight
recent progress.  Since RHIC has not operated yet, progress is confined
to construction, advances in carrying out realistic simulations, and
theoretical support.  I will emphasize what I view as the likely
flagship experiments with polarized beams at RHIC.

\section{The Tools}

\subsection{Acceleration and Monitoring of Polarized Beams}

The schematic layout of the RHIC accelerator complex, including the 
specific additions needed to produce colliding polarized beams, is shown
in Fig.\ 1.  The existing AGS polarized ion source will be replaced
before RHIC startup with an optically pumped source that was originally
built at KEK, and has now been substantially revitalized at TRIUMF.
The new source has already delivered 500 $\mu$A of pulsed beam in bench
tests at TRIUMF,\cite{Ze98} as will
be needed to attain the ``enhanced'' luminosities assumed later in this
talk: $8 \times 10^{31}$ cm$^{-2}$s$^{-1}$ for \pvec-\pvec\ collisions
at $\sqrt{s}$=200 GeV and $2 \times 10^{32}$ at $\sqrt{s}$=500 GeV.
With the partial 

\begin{wrapfigure}{l}{7.5cm}
\epsfig{figure=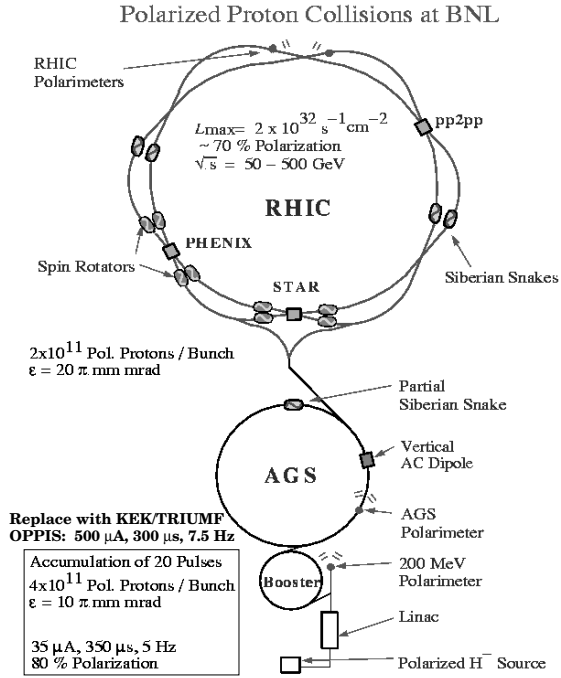,width=7.5cm}
\caption {\it 
Schematic layout of the RHIC accelerator complex, emphasizing the
devices important to the spin program.}
\end{wrapfigure}

\noindent Siberian Snake and rf dipole indicated in Fig.\ 1,
beam polarizations of 40--50\% have been attained to date in the AGS,
at energies suitable for injection into RHIC.  A plan to reach the 
desired 70\% has been developed, requiring installation of a new partial
helical snake in the AGS.

The first of the full superconducting helical di-
pole Snakes, intended to
allow acceleration of polarized beams through the many depolarizing
resonances in RHIC, was completed this past spring.  Installation by
summer 1999 of the two Snakes needed for one RHIC beam, together
with the first RHIC polarimeter on the same beam ring, will allow
commissioning of the first polarized proton beam in RHIC to begin during
its first year of operation.  Installation of the Snakes and polarimeters 
for the second beam is planned for the summer of 2000, along with the full 
complement of spin rotators needed
(see Fig.\ 1) to allow independent adjustment of the spin orientations
at the sites of the two major detectors.  It is then expected that the
first \pvec +\pvec\ collision studies will take place in year 2 of RHIC
operation.  In a conservative scenario, one may hope to reach ``design''
luminosity for polarized collisions in year 3, and the order-of-magnitude
enhanced luminosity
mentioned above by the start of year 4.  Together with various detector
upgrades described below, this schedule would permit the full RHIC
spin program to be carried out starting in 2002.  Present plans
call for polarized proton operation of RHIC for about 10 weeks per year,
beginning in Fall 2000.

The development of beam polarimeters for RHIC has been concentrated to
date on the polarized beam commissioning stage.  Polarimeters at first
will utilize fixed internal carbon micro-ribbon targets.  The primary
commissioning polarimeter will exploit the sizable analyzing powers
observed \cite{{AD91},{E925}} for inclusive production of charged pions at
moderate momentum transfer ($p_T$\grsim 1 GeV/c).  Funding shortages
limit the pion detection to a single-arm spectrometer assembled from
existing magnets.  A second option would utilize \pvec-carbon elastic
scattering in the Coulomb-nuclear interference region, where the low-energy
recoil carbon nuclei can be detected inexpensively in a left-right
symmetric setup,\cite{CNI} but analyzing powers are limited to a few
percent.  It is not clear that instrumental asymmetries can be kept 
sufficiently small ({\em i.e.,} nearly two orders of magnitude below
the polarization asymmetries) for either of these designs to be
adapted to the long-term needs of the RHIC spin program, where eventual
absolute calibration of beam polarizations to an accuracy of $\pm 5\%$
or better is needed.  For this and other reasons, a number of options
for long-term polarimetry are still under discussion.\cite{{St98},{Nu98}}  
The eventual absolute calibration of beam polarizations at
energies \grsim 100 GeV is likely to be performed against analyzing 
powers measured for the p-\pvec\ system with a polarized internal hydrogen
target, but details have yet to be worked out.

\subsection{The Major Detectors}

The bulk of the RHIC spin physics program will be carried out with the
two major RHIC detectors, PHENIX and STAR.  Both were originally
designed to analyze relativistic heavy-ion collisions, where the
performance demands are quite distinct from those for a p-p
collider.  However, both detectors are being adapted to provide 
cost-effective solutions for polarization measurements.
Assembly of both detectors in their respective experimental halls at RHIC is
actively under way at present, and parts of both will be used for
heavy-ion collision studies during the first year of RHIC operation.

\begin{figure}[bht]
\begin{minipage}[b]{7.5cm}
\begin{center}
\epsfig{figure=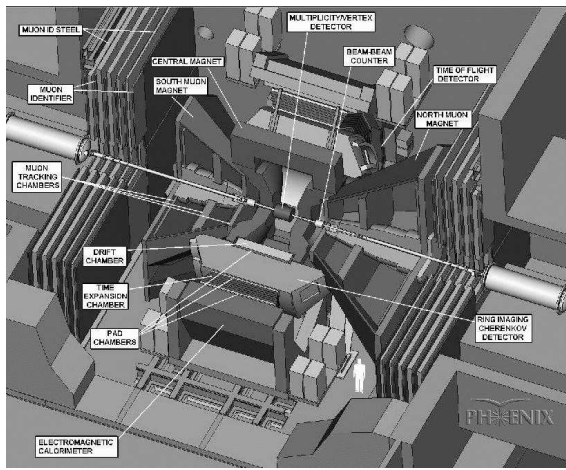,width=7.5cm}
\end{center}
\end{minipage}
\begin{minipage}[b]{4.2cm}
\begin{center}
\caption {\it 
Schematic of the PHENIX detector.  A second muon arm has been added
specifically as an upgrade aimed at the spin program.}
\end{center}
\vspace{20mm}
\end{minipage}
\end{figure}

As shown in Fig.\ 2, the PHENIX detector 
comprises central arms and fore and aft muon detectors.  The central
arms provide high-resolution charged-particle tracking and electromagnetic
calorimetry, but over a limited solid angle: pseudorapidities $|\eta| \leq
0.35$ and about half of the full azimuthal range.  The muon arms cover
the pseudorapidity range $1.2 \leq |\eta| \leq 2.4$.  The primary upgrade
of PHENIX aimed specifically at the spin program is the addition of the
second muon arm.

The STAR detector (Fig.\ 3) is intended to provide a more global view of
each collision.  Time projection chambers are used for high-resolution
tracking of many charged particles over the full azimuthal range and
most of the pseudorapidity range $|\eta| \leq 3$.  The major upgrades
relevant to the spin program are electromagetic calorimeters (EMC) providing
full azimuthal coverage over 
the barrel portion of the cylindrical surface ($|\eta| \leq 1.0$) and
one endcap ($1.07 \leq \eta \leq 2.0$).  Construction of the barrel EMC 
is under way, while a funding proposal for the endcap is in preparation.  
With suitable funding profiles, both EMC's can be completed by the summer
of 2002.

\begin{figure}[thb]
\begin{minipage}[b]{8.5cm}
\begin{center}
\epsfig{figure=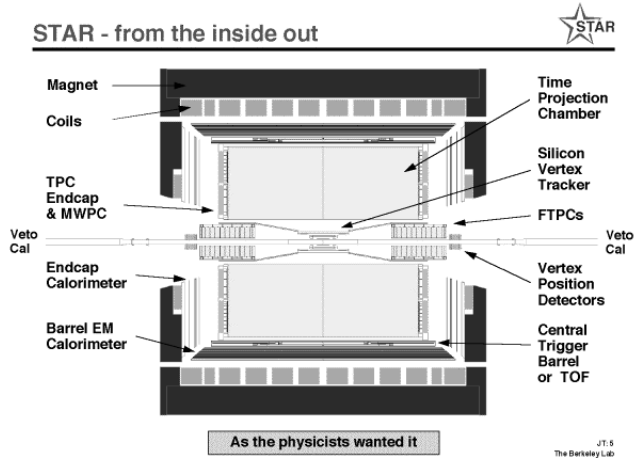,width=8.5cm}
\end{center}
\end{minipage}
\begin{minipage}[b]{3.2cm}
\begin{center}
\caption {\it Cross-section view of the subsystems of the cylindrically
symmetric STAR detector.  Construction has begun on the barrel portion
of the electromagnetic calorimeter (EMC), while funding is still being
sought for one endcap EMC. }
\end{center}
\vspace{4mm}
\end{minipage}
\end{figure}

The two detectors are complementary in their suitability to the proposed
spin program.  PHENIX has a distinct edge in hadron and muon particle 
identification and in rate capability.  The muon arms considerably
facilitate studies of the production of dilepton pairs, W$^\pm$ and
J/$\Psi$.  The fine-grained EMC in PHENIX provides very good capability
to distinguish photons from $\pi^0$'s over the energy range of interest
near mid-rapidity.  On the other hand, the expanded $\eta$ coverage
available in STAR for charged particles and photons makes it the detector
of choice for hadron jets and photon-jet coincidences.  The broad
coverage also considerably extends the range of
Bjorken $x$-values that can be accessed for colliding partons,
and facilitates imposition of isolation cuts to distinguish lone
particles from jet fragments.  Although the EMC's in STAR have generally much
coarser segmentation than those in PHENIX, they include fine-grained
detectors at the depth of maximum electromagnetic shower energy deposition,
to aid in distinguishing between the shower profiles characteristic of 
single high-energy photons {\em vs.} daughter di-photons from neutral meson 
decays.  The spin physics goals of the two collaborations are similar,
but both detectors are needed to address these goals optimally.

\subsection{Theoretical Tools}

The final crucial tool for the RHIC spin physics program is perturbative
QCD.  The underlying assumption of the entire program is that the
description of hard
($p_T$\grsim 10 GeV/c) proton collisions at RHIC energies can be
factorized into a non-perturbative structure part, about which we seek
to learn, and perturbative partonic collisions. 
The colliding proton beams are thus viewed
as ensembles of polarized parton beams; the parton luminosities and
polarizations are related to those of the protons by the
spin-independent and spin-dependent parton distribution functions (PDF's) 
that characterize the proton's substructure.  
The parton-level cross sections and spin observables are given
straightforwardly by leading-order (LO) pQCD \cite{BS89} or, in the case
of electroweak (e.g., W$^\pm$ production) processes, by the Standard 
Model.  In
fact, next-leading-order (NLO) calculations have already been performed
for many of the processes central to the RHIC spin program: they provide
important quantitative corrections, but generally do not alter the
qualitative design of experiments.  Nonetheless, thorny issues remain,
associated, for example, with potential spin effects arising from
multiple soft gluon radiation\cite{Sterman} preceding hard partonic
collisions.  A large and very active group of theorists are addressing
these issues and present indications remain very encouraging for the
clean interpretability of RHIC spin results.  But it will clearly be wise
in the early years of RHIC to test pQCD predictions for spin effects, 
{\em e.g.}, by checking that single-spin transverse asymmetries involving
light quarks really do approach zero \cite{BL81} at sufficiently high $p_T$, 
and by comparing spin structure results from proton collisions to those 
obtained with electromagnetic probes.

\section{Snapshots of the RHIC Spin Physics Program}

\subsection{Determination of the Gluon Helicity Distribution}

As we have already heard in several previous talks at this conference,
the {\em next essential piece} of the nucleon spin puzzle is measurement
of the gluon helicity preference $\Delta G(x,Q^2)$:  the difference between
the PDF's for gluons with spin projection along {\em vs.} opposite the 
nucleon's overall longitudinal spin projection.  It is already known that
gluons dominate the {\em mass} of the proton.\cite{SH78}  Do they also
make crucial contributions to the spin?

By virtue of the axial anomaly of QCD,\cite{CH96} there are contributions
from the integral $\Delta G(Q^2) \equiv \int_0^1 {\Delta G(x,Q^2) dx}$, 
as well as from quark
and antiquark helicity preferences, to the integrated asymmetry measured
in polarized deep inelastic scattering (DIS) experiments.  Thus, these
experiments strictly determine only a correlation between the integrated
quark and gluon helicity preferences, as illustrated by the SMC analysis
\cite{AD97} in Fig.\ 4.  An independent measurement of $\Delta G$ is needed
even to pin down the {\em quark} contributions to the proton spin.  An 
experimental precision better than $\pm 0.5$ in $\Delta G$ is required to
reduce the ambiguity in quark contributions to a level comparable to the
actual measurement errors in the DIS asymmetries.  The quark and gluon 
contributions must be determined
separately before we can constrain the fraction of the nucleon spin 
attributable to {\em orbital} angular momentum of the partons.

\begin{figure}[bht]
\begin{minipage}[b]{5.0cm}
\begin{center}
\epsfig{figure=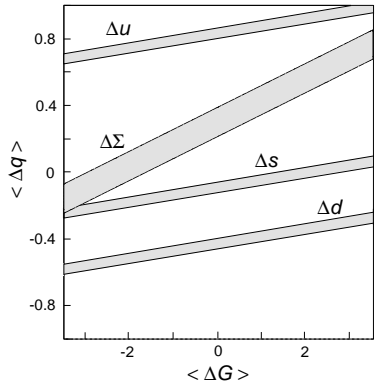,width=5cm}
\end{center}
\end{minipage}
\begin{minipage}[b]{6.5cm}
\begin{center}
\caption {{\it The correlation of quark and gluon contributions to the
longitudinal polarization of a proton, introduced by the effect of
the QCD axial anomaly on the interpretation of polarized DIS
asymmetries.  The separation of quark and
gluon contributions to polarized DIS is scheme-dependent, and is shown
here for one particular QCD factorization scheme.  The bands in each case
represent $\pm 1\sigma$ limits on the quark contributions deduced  
from DIS results in Ref.~14, from which the figure is taken.}}
\vspace{8mm}
\end{center}
\end{minipage}
\end{figure}

At present there are only coarse constraints on $\Delta G(x,Q^2)$ available
from observed scaling violations in polarized DIS.  The three
quite different distributions for $\Delta G(x,Q^2)$ shown in Fig.\ 5, resulting
from NLO fits of the DIS data by Gehrmann and Stirling,\cite{GS96} 
give some indication of the range of values consistent with the data.  In
all models for the gluon spin distribution suggested to date, the dominant
contributions to the integral $\Delta G(Q^2)$ arise from Bjorken $x$-values
below 0.1, simply because that is where most of the gluons reside.  RHIC
spin experiments can probe the range 0.01\lesim $x_{gluon}$\lesim 0.3, 
sufficient to deduce the integral with the desired precision.

\begin{figure}[thb]
\begin{minipage}[b]{6.0cm}
\begin{center}
\epsfig{figure=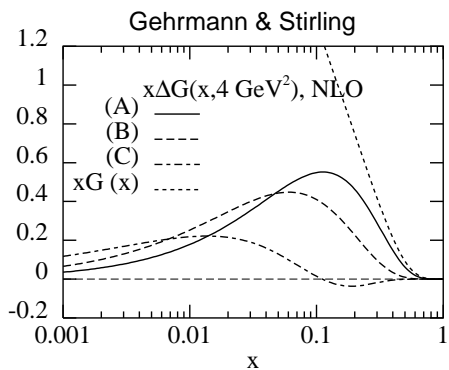,width=6cm}
\end{center}
\end{minipage}
\begin{minipage}[b]{5.5cm}
\begin{center}
\caption {{\it Three models of the gluon helicity distribution used
in next-leading-order global analyses of polarized DIS results by 
Gehrmann and Stirling (Ref.~15).  All three are consistent with observed 
scaling violations in DIS.  The solid curve (A), corresponding to an 
integral $\Delta G(4$~GeV$^2$)=1.71, is used in simulations
presented here.}}
\end{center}
\vspace{8mm}
\end{minipage}
\end{figure}

Many processes in \pvec +\pvec\ collisions offer sensitivity
to $\Delta G(x,Q^2)$: {\em e.g.,} direct photon production via gluon Compton
scattering ($q+g \rightarrow q + \gamma$); dijet or di-hadron production via
quark-gluon or gluon-gluon elastic scattering; open heavy quark ($c$ or $b$)
production via gluon-gluon fusion.  All of these examples will be measured
by the PHENIX and/or STAR collaborations, and the results from several such
processes are likely to be included in eventual NLO analyses of polarization
data to extract $\Delta G(x,Q^2)$.  However, one should keep in mind that
such analyses are subject to many intertwined theoretical ambiguities:
competing LO partonic processes contributing to the measured channels; 
non-negligible NLO corrections; several relevant, and poorly delineated,
polarized PDF's; pQCD scale ambiguities; uncertain treatments of experimental
isolation cuts and of poorly known aspects of fragmentation functions; the
effects of multiple soft gluon radiation, and of the transverse initial
parton momentum components (``$k_T$-smearing'') they introduce.\cite{HU95}
In light of these ambiguities, {\em the best constraints on the extracted
polarized gluon PDF should be expected from those experiments that
approach most closely the ideal of a direct LO measurement of the gluon 
helicity preference at experimentally determined values of} $x_{gluon}$. 
Based on this criterion, I concentrate below on one
particular RHIC experiment:  the measurement of the longitudinal spin
correlation $A_{LL}$ for \pvec\ + \pvec\ $\rightarrow \gamma$ + 
jet + X with the STAR detector.

The advantages of direct photon production and of $\gamma$-jet coincidence
detection can be summarized as follows:

\begin{itemize}
\item[1)] There is a single dominant LO pQCD process: $q + g \rightarrow q +
\gamma$.  The main LO background, from annihilation $q + \overline{q} 
\rightarrow \gamma + g$, contributes at the $\sim 10\%$ level.  NLO 
calculations
have been performed,\cite{GV94} and indicate no qualitative changes from the
LO expectations (except for a slightly {\em enhanced} sensitivity to 
$\Delta G$).  Higher-twist corrections are expected to remain negligible 
at $p_T$\grsim 10 GeV/c.

\item[2)] The sensitivity to gluon polarization is guaranteed to be large
in an experiment with appropriate kinematic coverage.  The pQCD spin 
correlation
for gluon Compton scattering approaches unity when the $\gamma$ is detected
in the direction of the incident quark (where the cross section for the
process is also maximized).  Large quark polarizations (\grsim 30\%) are
available at momentum fractions $x_{quark}$\grsim 0.2 to probe $\Delta G$.
It is then highly desirable to sample very asymmetric partonic collisions
($x_q \geq 0.2$ with $x_g \leq 0.1$), in which both products will be boosted
forward in the lab frame.  Coverage for such asymmetric collisions requires
the endcap EMC proposed for STAR, which then also spans the appropriate range
of partonic c.m. angles, where the spin correlation is large.

\item[3)] Detection of $\gamma$-jet coincidences allows event-by-event 
kinematic reconstruction of the momentum fractions $x_{1,2}$ for the colliding 
partons.  It is this coincidence detection involving jets that requires the
large acceptance of STAR, and that facilitates the {\em direct} extraction of
$\Delta G(x,Q^2)$ from the data.  The combination of coincidence and
polarization measurements also turns out to yield considerably reduced
sensitivity (in comparison to cross section measurements aimed at determining
the {\em unpolarized} gluon PDF) to the kinematic uncertainties arising from 
$k_T$-smearing.\cite{PAC}

\item[4)] Measurements with STAR will cover a suitably broad range of
momentum fractions, 0.01\lesim $x_g$\lesim 0.3, to determine the integral
contribution of gluons to the proton helicity with a precision better than
$\pm 0.5$.  This coverage requires runs at two bombarding energies, 
$\sqrt{s}=200$ and 500 GeV.  Adequate statistical precision can be obtained
with 10-week runs at each energy, at ``enhanced'' \pvec +\pvec\ luminosities.
In an optimistic scenario, these runs could take place in 2002-3. 

\end{itemize}

\begin{figure}[thb]
\begin{minipage}[b]{\textwidth}
\begin{center}
\epsfig{figure=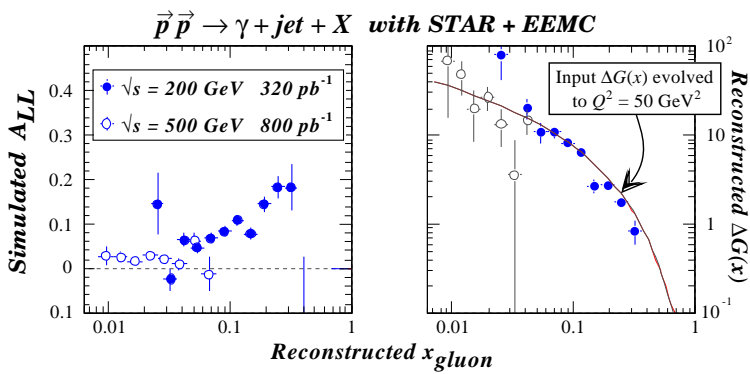,width=10cm}
\end{center}
\end{minipage}
\vspace{6mm}
\begin{minipage}[b]{\textwidth}
\caption {{\it Simulation results for the pp spin correlation $A_{LL}$
and the gluon helicity distribution $\Delta G(x)$ extracted therefrom,
for photon-jet coincidence events detected in STAR (including its planned
endcap EMC) at $\sqrt{s}$=200 GeV (closed symbols) and 500 GeV (open
symbols).  The events analyzed in Ref.~18 
have been subjected to cuts foreseen for the real data.  The error
bars reflect counting statistics for 10-week runs at each energy, assuming
${\vec p} + {\vec p}$ luminosities $\sim 10^{32}$ cm$^{-2}$s$^{-1}$.
The solid curve in the right-hand frame represents the theoretical input for
$\Delta G(x,Q^2=50$(GeV/c)$^2$).  The small systematic deviations between the
input and extracted gluon helicity distributions arise from simplifying 
assumptions in the data analysis, and are correctable via simulations, as
discussed in the text.}}
\end{minipage}
\end{figure}

Simulation results that demonstrate the quality of data attainable for
the extracted $\Delta G(x)$ are shown in Fig.\ 6.  Events were generated
with the code PYTHIA,\cite{SJ94} version 5.7, incorporating all LO
photon production processes, plus initial-state gluon radiation and splitting
that give rise to $k_T$-smearing, plus final-state parton fragmentation.  
LO spin effects for all the hard partonic
processes were included \cite{PAC} as appropriate for proton beam
polarizations of 0.7 and Gehrmann-Stirling spin-dependent PDF's \cite{GS96}
evolved with $Q^2$ (chosen equal to $p_T^2/2$).  The gluon helicity 
distribution used as
input to the simulations was set A from Ref.\ \cite{GS96}, as
represented by the solid curve in Fig.\ 6, 
after evolution to a value of $Q^2$ corresponding to the minimum -- and 
most probable -- transverse momentum transfer included in the simulations,
$p_T$=10 GeV/c.  Although $\Delta G(x,Q^2)$ continues to rise rapidly
down to the lowest $x_g$ values covered in the proposed experiment, the
gluon {\em polarization} $\Delta G(x)/G(x)$ decreases with decreasing $x_g$,
producing the decrease in simulated $A_{LL}$ values seen in the left-hand
frame.  

The projected results for $\Delta G(x)$ shown in the right-hand frame
of Fig.\ 6 were extracted from the simulated asymmetries with a 
simple-minded reconstruction algorithm, which assumed 
that {\em only} the gluon Compton scattering process
contributed to the events, and then always with $k_T = 0$ and with
$x_g = {\rm min}[x_1,x_2]$.  The neglect of other contributions included
in the generated events leads to the small systematic deviations of the
extracted $\Delta G(x)$ from the input curve at the larger $x_g$-values
seen in Fig.\ 6.  These
deviations can eventually be compensated in an experiment by using
simulations to introduce appropriate corrections, but in any case they 
have little effect on the integral gluon spin contribution deduced from
the results.  A {\em fit} to the data points in the right-hand frame with the
functional form used for $\Delta G(x)$ by Gehrmann and Stirling \cite{GS96}
gives an extracted integral $\Delta G_{recon}$ = 1.62 $\pm$ 0.23. 
The error bar on the fitted value,
which includes the uncertainty in extrapolating the data to $x_g = 0$, is
reduced by a factor of 6 by the inclusion of the 500 GeV data.  It does not
include other systematic errors, associated, for example, with subtraction
of background from $\pi^0$ and $\eta^0$ production and with beam polarization
absolute calibration uncertainties.  When such other systematic errors are
included the error bar on the integral is likely to increase to about
$\pm 0.4$.  The quality of these projected results is thus comparable to
that contemplated with an ${\vec {\rm e}}$-\pvec\ collider at HERA\cite{RA97}
and is distinctly superior to that of other measurements planned for the 
coming decade.

\subsection{Antiquark Polarizations from W$^\pm$ Production}

Polarized DIS experiments have suggested that the sea in a polarized
proton is significantly polarized.\cite{AD97}  It is important for our
understanding of the structure of the sea to know if this polarization
is shared by antiquarks as well as quarks, and if it is flavor-dependent.
The study of intermediate vector boson production in RHIC \pvec +\pvec\
collisions provides a very clean way to address these questions.  Since
the production processes are weak, one can measure parity-violating
{\em single} spin asymmetries $A_L^{PV}$ associated with spin-flip of
{\em each} of the two polarized proton beams (referred to below as beams
$a$ and $b$).  In kinematic regimes ($|\eta| > 1$) sampling quite
asymmetric parton collisions ({\em e.g.,} $x_a >> x_b$), this allows one
to separate the contributions from polarized (predominantly valence) quarks
{\em vs.} polarized antiquarks.  

This separation is most easily illustrated for W$^\pm$ production, where
there is, in each case, a single dominant partonic process:  $u + 
\overline{d} \rightarrow {\rm W}^+; ~ d + \overline{u} \rightarrow {\rm W}^-$.
The parity-violating analyzing powers can be written straightforwardly in
terms of the relevant PDF's; \cite{BS93} {\em e.g.,}
\begin{equation}
A_L^{PV}({\rm W}^+, {\rm beam}~a) ~=~ {\Delta u(x_a) \overline{d}(x_b)
- \Delta \overline{d}(x_a) u(x_b) \over u(x_a) \overline{d}(x_b) +
\overline{d}(x_a) u(x_b) + hfc},
\label{eq:alw}
\end{equation}
\noindent where $hfc$ refers to small contributions from heavier quark 
flavors.  In the limit $x_a >> x_b$, these general results tend toward
direct measurements of the quark and antiquark polarizations:
\begin{equation}
x_a >> x_b \Rightarrow A_L^{PV}({\rm W}^+, {\rm beam}~a) \rightarrow 
{\Delta u \over u}(x_a); ~~A_L^{PV}({\rm W}^+, {\rm beam}~b) 
\rightarrow {\Delta \overline{d} \over \overline{d}}(x_b);
\end{equation} 
\begin{equation}
x_a >> x_b \Rightarrow A_L^{PV}({\rm W}^-, {\rm beam}~a) \rightarrow 
{\Delta d \over d}(x_a); ~~A_L^{PV}({\rm W}^-, {\rm beam}~b) 
\rightarrow {\Delta \overline{u} \over \overline{u}}(x_b). 
\end{equation}
\noindent One thus has, in principle, the capability to calibrate the extraction
of antiquark polarizations by comparing the quark polarizations measured
simultaneously (via spin-flip of the {\em other} beam) to results already
known from DIS.  This calibration is crucial for the entire RHIC spin
program, because it represents one of the few points of direct confrontation
of results obtained via hadronic probes with those from electromagnetic probes.

Figure 7 illustrates the range of $x$-values over which quark and antiquark
polarizations could be sampled, and the statistical sensitivity attainable
for \pvec +\pvec $\rightarrow {\rm W}^\pm$+X processes measured at
$\sqrt{s} = 500$ GeV with the PHENIX detector.  (STAR will measure
these processes with comparable sensitivity.)
The predicted analyzing powers \cite{BS93} are large and
strongly sensitive to the antiquark polarizations over the range 0.05\lesim
$x_{\overline{q}}$\lesim 0.1.  
The separation of quark from antiquark contributions is not quite so clean
as suggested by Fig.\ 7.  The W's are detected in practice via single, 
isolated 
hard leptons (e$^\pm$ or $\mu^\pm$ at $p_T$\grsim 25 GeV/c) from their 
decay, and this complicates the identification of events where $x_a >> x_b$.
Such events completely dominate the W$^-$ sample at $|\eta|_e > 1.0$,
but the W$^+$ sample always contains sizable contributions from both
$u_a + \overline{d}_b \rightarrow {\rm W}^+$ and 
$u_b + \overline{d}_a \rightarrow {\rm W}^+$ processes.\cite{PAC}  The 
difference
arises between the two cases because both W$^\pm$ are produced left-handed,
so that the daughter leptons are emitted preferentially {\em along} the
W$^-$ momentum direction, but {\em opposite} the W$^+$ direction.  Thus,
it is the measurement of down quark polarizations from W$^-$ production
that will afford the best comparison of RHIC results with those from
(semi-inclusive) polarized DIS.

\subsection{Theoretical Predictions for Other RHIC Spin Measurements}

In addition to the very substantial recent progress made on machine and 
detector design and construction, as well as on experiment simulations, 
for the RHIC spin program, there has been a steady flow of new ideas for
experiments supplied by interested theorists.  I briefly review a few 
recent highlights here.

\begin{wrapfigure}{l}{5.5cm}
\epsfig{figure=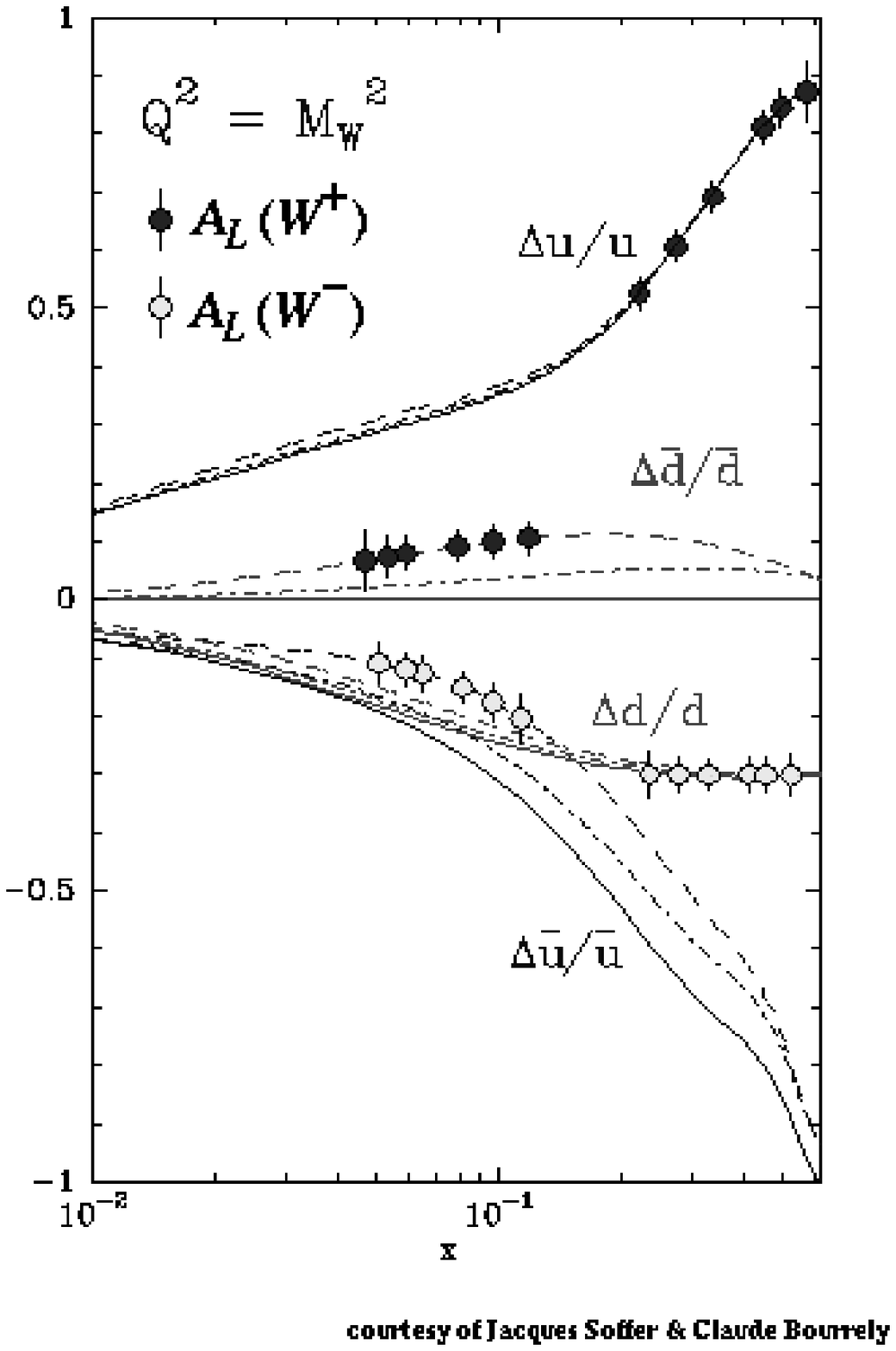,width=5.5cm}
\caption {{\it The Bjorken $x$ ranges and statistical precision goals
for PHENIX measurements of $u$ and $d$ quark and antiquark polarizations,
via single-spin parity-violating asymmetries for inclusive $W^\pm$ 
production at $\sqrt{s} = 500$ GeV.  Only those
events in which a daughter muon from the $W$ decay is detected in one of
the muon arms are included.  The curves represent various model predictions
from Ref.~21 for the spin-dependent structure functions.}}
\end{wrapfigure}

There continues to be strong interest in the distributions of
{\em transversely} polarized quarks in a transversely polarized proton.
These transversity distributions $\Delta_T q(x,Q^2)$ are expected to
differ from the helicity distributions $\Delta q(x,Q^2)$ by virtue of
relativistic quark behavior, and they provide new and independent 
information about nucleon spin structure.\cite{JA97}  In particular, 
when expressed
in a helicity basis, the transversity measures the chiral-odd probability
that a polarized nucleon may undergo a fluctuation in which it emits a
quark of one helicity, and then absorbs a quark of the opposite 
helicity.\cite{JA97}  Transversity can be measured, in principle, via transverse
spin correlations ($A_{TT}$) in hard \pvec-\pvec\ collisions.  Drell-Yan
dilepton or Z$^0$ production represents a cleanly interpretable case,
except for the fact that unknown quark transversities will be folded
there with unknown, and perhaps very small, antiquark transversities.
Martin {\em et al.}\cite{Ma98} have recently estimated the largest
effects that might be expected from transversities in Drell-Yan 
production, by exploiting an inequality first suggested by Soffer \cite{SO95}: 
$2 |\Delta_T q(x,Q^2) | \leq q(x,Q^2) + \Delta q(x,Q^2).$
The predicted effects are small, but measurable.  For example, at
$\sqrt{s}=150$ GeV, they predict maximal $A_{TT}$ values of plus several
percent in the dilepton mass range from 10--20 GeV, to be compared
to the small negative values expected in the absence of relativistic
effects ({\em i.e.,} for $\Delta_T q$ = $\Delta q$).\cite{Ma98}
Alternative reactions for exploring transversity, without the sensitivity
to antiquarks, are also under active consideration.\cite{JA97}

Tests of spin substructure models can be extended, in principle, beyond
nucleons to hyperons, if one can measure the polarization of
such hadrons (via their weak decay) when they appear as substantial 
fragments of a partonic jet.
Measurements of this sort by the OPAL \cite{OPAL} and ALEPH \cite{ALEPH}
collaborations, for ${\rm e}^+{\rm e}^-$ collisions at the Z$^0$ resonance,
seem consistent with the simplest model of $\Lambda$ spin structure,
in which the spin is carried completely by the strange valence quark.  
However, de Florian {\em et al.} have pointed out \cite{DeF98} that the
contrast with the complexity of the nucleon spin structure is not 
necessarily so striking, since the LEP results can also accommodate quite
different models of the $\Lambda$'s spin/flavor structure.  As 
illustrated in Fig.\ 8, these models
could be very clearly distinguished by RHIC measurements of longitudinal
polarization transfer ($D_{LL}$) in inclusive $\Lambda$ production:
\pvec +p$\rightarrow {\vec \Lambda}$+X.  The inferred polarized fragmentation
functions will be sensitive not only to the $\Lambda$ structure, but also
to its parentage:  one should expect quite different results for $\Lambda$'s
that are direct fragments and for decay daughters of heavier hyperons
({\em e.g.,} ${\vec \Sigma}^0 \rightarrow
{\vec \Lambda} \gamma$ or ${\vec \Xi}^0 \rightarrow {\vec \Lambda} \pi^0$)
with very different internal spin coupling.  From this viewpoint, it would 
be desirable to detect
photons or $\pi^0$ correlated with the $\Lambda$, along with its daughter
proton and $\pi^-$, requiring a large acceptance detector with good 
charged-particle tracking and electromagnetic calorimetry.  STAR appears
to fill this bill, but relevant simulations have yet to be performed.

\begin{figure}[bht]
\begin{minipage}[b]{6.5cm}
\begin{center}
\epsfig{figure=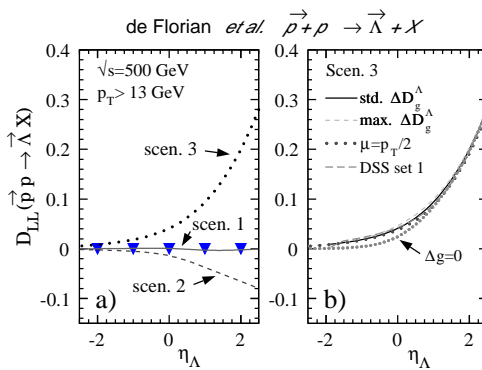,width=6.5cm}
\end{center}
\end{minipage}
\begin{minipage}[b]{5.2cm}
\begin{center}
\caption {{\it Predictions from Ref.~27 for helicity transfer
from a polarized proton beam to a $\Lambda$ jet fragment, based on 
various model scenarios of the $\Lambda$ spin structure and the
fragmentation process, all consistent with polarized fragmentation
measurements at LEP.  RHIC measurements for $\Lambda$'s produced
at pseudorapidity $\eta$\grsim 1 could easily distinguish among the
scenarios.}}
\end{center}
\vspace{4mm}
\end{minipage}
\end{figure}

An exciting possibility at RHIC is to use parity violation in hard jet
production to probe potential new interactions of very short range, beyond
the Standard Model.  Parity violation does arise within
the Standard Model for quark-quark scattering, from interference between
gluon- and Z$^0$-exchange.  

\begin{wrapfigure}{l}{6.0cm}
\epsfig{figure=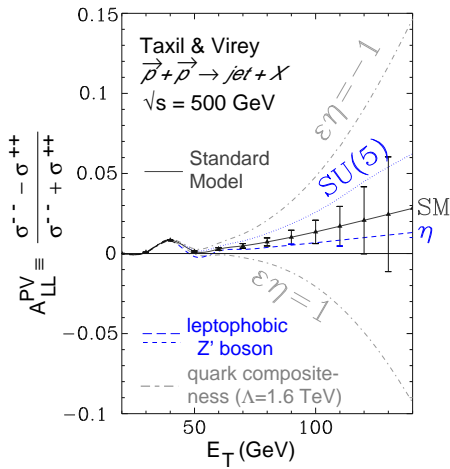,width=6.0cm}
\caption {{\it Predictions from Ref.~28 for two-spin parity-violating
asymmetries in hard jet production, based on mechanisms within and 
beyond the Standard Model.  The error bars reflect estimates of the
statistical uncertainties attainable with STAR in one standard year
of ${\vec p} + {\vec p}$ running.}}
\end{wrapfigure}

\noindent Predictions \cite{TV95} for the resulting two-spin asymmetries (measuring
sensitivity of the cross section to the simultaneous flip of both beam
helicities) for jet production at $p_T \sim 100$ GeV/c are shown in
Fig.\ 9.  Such hard collisions may also be sensitive to 
interference with amplitudes associated with new phenomena at a mass scale
\grsim 1 TeV.  The calculations in Fig.\ 9 include consideration \cite{TV95} 
of two
such classes of phenomena, associated either with quark compositeness
or with a new heavy ``leptophobic'' Z$^\prime$ boson.  The present limits
on such new phenomena still allow modifications to the parity-violating
asymmetries that are large in comparison to the uncertainties contributed
by PDF errors to the Standard Model predictions.  With realistic
measurement uncertainties, RHIC \pvec +\pvec\ experiments should attain
sensitivities at $\sqrt{s}=500$ GeV comparable to those that can be reached 
in 2 TeV unpolarized $\overline{{\rm p}}$p collisions at the 
Tevatron.\cite{TV95}

\section{Conclusions}

There can no longer be any doubt that a productive program of polarization
measurements in \pvec +\pvec\ collisions at several hundred GeV c.m.
energies will take place at RHIC in the first years of the next century.
Strong spin subgroups have formed within both the PHENIX and STAR
collaborations.  Upgrades to facilitate the spin research program are
well under way for the accelerator complex and for both major detectors.
Though many details remain to be addressed, I envision forefront
research with this polarized collider for at least a 10-year period, with
emphasis on: (1) the best measurement accessible in the coming decade of 
the gluon contribution to the proton spin; (2) separation of
flavor-dependent antiquark helicity preferences within a polarized
proton, via W$^\pm$ production; and (3) searches for new ultrashort-ranged
parity-violating interactions in hard quark-quark scattering at 
$p_T \sim 100$ GeV/c.  Prospects for other groundbreaking results are
very encouraging:  theorists are proposing new ideas for
RHIC spin experiments faster than we can simulate them.  Very recent
examples include suggestions to determine
polarized fragmentation functions in $\Lambda$ production \cite{DeF98}
and to search for intrinsic charm in the proton.\cite{Be98}

The RHIC spin program will blossom fully after \pvec +\pvec\ luminosities
$\approx 10^{32}$ cm$^{-2}$s$^{-1}$ are reached.  But even before that, in
the second and third years of RHIC operation, PHENIX and STAR will gather
important information on polarization systematics in hadronic processes
and first indications of the gluonic contribution to the proton spin.
Strong spin physics is also accessible beyond STAR and PHENIX, including
a proposed experiment \cite{pp2pp} to map out polarization observables for 
\pvec-\pvec\ elastic scattering over a broad range of momentum transfers,
at unprecedentedly high energies.

In short, as RHIC spin rapidly approaches the ``real axis,'' there
is a great deal of work to do!

\end{document}